\documentstyle[aps,prd]{revtex} 

\begin{document}
\title{Mode decomposition and renormalization in semiclassical gravity}
\author{Albert Roura and Enric Verdaguer
\thanks{also at Institut de F\'\i sica d'Altes Energies (IFAE)}
}
\address{Departament de F\'{\i}sica Fonamental,
Universitat de Barcelona, Av.~Diagonal 647,\\
08028 Barcelona, Spain}
\date{\today}
\maketitle
\draft

\begin{abstract}
We compute the influence action for a system
perturbatively coupled to a linear scalar field acting as the environment.
Subtleties related to divergences that appear when summing over all the
modes are made explicit and clarified. Being closely connected with models
used in the literature, we show how to completely reconcile the results
obtained in the context of stochastic semiclassical gravity when using
mode decomposition with those obtained by other standard functional
techniques.
\end{abstract}

 
\section{INTRODUCTION} 
 
The Closed Time Path (CTP) functional formalism has been very useful to
study the back-reaction effect in the context of semiclassical gravity (as
well as other aspects of field theory)
\cite{jordan86/calzetta87/campos94,hu93,calzetta95,calzetta97}.
When considering the back-reaction problem in semiclassical gravity, one
is usually interested only in the gravitational field dynamics whereas the
quantum matter fields are treated as an environment
\cite{calzetta94,campos96}. The results obtained when integrating out the
environment degrees of freedom are closely connected with the influence
functional \cite{feynman}, a statistical field theory method which has
proved very fruitful to reveal the stochastic nature of open quantum
systems (for applications to quantum Brownian motion models see Ref.
\cite{caldeira83/hu92/gellmann93}). In fact, it has been pointed out that
semiclassical gravity \cite{hu89,calzetta94} and effective theories in
general should exhibit dissipation and noise \cite{calzetta97}.
To describe the stochastic character of the system dynamics due to the
noise induced by the environment, Langevin-type equations are required.
Thus, Einstein-Langevin equations have been used to address the
back-reaction problem in the framework of semiclassical gravity
\cite{calzetta94,hu95,campos96,charo99}.
 
When dealing with fields, mode decomposition can be a useful calculational 
tool since it makes the problem closer to quantum mechanical systems (free
fields are treated as an infinite set of decoupled harmonic oscillators).
The main advantage of this method is that the noise and dissipation
kernels can be obtained in a rather direct way \cite{hu94} and, in the
context of semiclassical gravity, provides a simple connection with the
Bogoliubov coefficients (closely related to particle creation effects)
\cite{calzetta94,hu95}. For each mode no renormalization is required, the
need for renormalization arises when considering an infinite number of
degrees of freedom: it is precisely when summing over all the modes that
one gets ultraviolet divergences.
However, the appearance of distributional functions makes this sum rather 
subtle, the presence of such divergences is not always manifestly evident 
and misleading results may be obtained.
In the semiclassical gravity context this is particularly important as one
may overlook the need for counterterms to renormalize the divergences,
which will imply the appearance of finite extra terms when addressing the
back-reaction problem. 
These drawbacks do not arise in other treatments based on functional
methods typical of quantum field theory which make no use of mode
decomposition, where renormalization seems to be more easily handled
\cite{jordan86/calzetta87/campos94,campos96,calzetta98}. 

The aim of this report is to show how to reconcile the results obtained by 
means of a mode-decomposition approach with the results based on standard 
field theory techniques for renormalization in curved space-times
\cite{calzetta98}.
In Sect. \ref{sect2} we introduce the notation and the model that we are
going to work with and evaluate the influence action perturbatively.
A concrete example is considered in Sect. \ref{sect3}, where sum over
modes is performed revealing the appearance of divergences, and it is
shown how they can be handled. In Sect. \ref{sect4}, the previous results
are used
to consider models treated in the literature which use mode decomposition
in the context of stochastic semiclassical gravity and show how to
reconcile these results with those obtained by usual functional methods.
 
\section{MODE-DECOMPOSED EXPRESSION FOR THE INFLUENCE ACTION} 
\label{sect2}

To make the description as simple as possible we follow ref.
\cite{calzetta94} and consider the whole action for a system described
by the variable $x(t)$ with action $S[x(t)]$ and the environment,
described by a free field $\phi(t,\vec x)$ in flat space which has been
decomposed in a complete set of modes $\{u_k(\vec x)\}$: $\phi(t,\vec x)=
\sum_k q_k(t) u_k(\vec x)$. The free action
$S[\phi(t,\vec x)]$ is local and at most quadratic in $\phi(t,\vec x)$,
since the field is linear. If that is also the case for the term
$S^{int}[x(t),\phi(t,\vec x)]$ describing the interaction with the system,
the action terms for the environment may be written, after performing the
spatial integrals and using the completeness relation for the modes, as
$\sum_k S[q_k(t)]$ and $\sum_k S^{int}_k[x(t),q_k(t)]$
respectively, where the action for each mode $S[q_k(t)]$ corresponds to
that of a harmonic oscillator. The dynamics is therefore equivalent to
that of a set of decoupled harmonic oscillators interacting separately
with the system.
\begin{eqnarray} 
\label{1} 
S[x(t),\phi(t,\vec x)] =S[x(t)] +\sum\limits_k S[q_k(t)] 
+\sum\limits_k S^{int}_k[x(t),q_k(t)] ,  
\end{eqnarray}
where $S^{int}_k[x(t),q_k(t)] =\int dtQ _k(q_k(t))\cdot h(x(t))$ with
$Q_k$ and $h$ being some specific functions. 
The expression for the Feynman and Vernon influence functional
\cite{feynman} in the interaction picture is:  
\begin{eqnarray} 
\label{2}F\left[ x_{+},x_{-}\right] =e^{iS_{IF}\left[ x_{+},x_{-}\right] 
}=\,_I\left\langle 0\ \rm{in}\right| \prod\limits_k T^-\left( 
e^{-iS^{int}_k[x_{-}(t)]}\right) T^+\left( e^{iS^{int}_k[x_{-}(t)]}\right) 
\left| 0\ \rm{in}\right\rangle _I  ,
\end{eqnarray} 
where $T^+$ and $T^-$ correspond to the time ordering and
anti-time ordering prescriptions respectively. To obtain the influence
action $S_{IF}$, we will treat the interaction term $S^{int}_k[x(t)]$
perturbatively. Taking the logarithm of (\ref{2}) and expanding up to
second order in $S^{int}_k$, we get: 
 
\begin{eqnarray} 
\label{3} 
S_{IF}[x_{+},x_{-}] &\simeq& \int dt [G_{+}(t)x_{+}(t)+G_{-}(t)x_{-}(t)] +  
\frac 12\int dtdt^{\prime}[G_{++}(t,t^{\prime})x_{+}(t)x_{+}(t^{\prime}) +
G_{+-}(t,t^{\prime})x_{+}(t)x_{-}(t^{\prime}) \nonumber \\ 
&& +G_{-+}(t,t^{\prime})x_{-}(t)x_{+}(t^{\prime})+G_{--}(t,t^{\prime}) 
x_{-}(t)x_{-}(t^{\prime})] ,
\end{eqnarray} 
where $G_{+}(t) = -G_{-}(t) =\sum_k\langle Q_k(q_k(t)) \rangle$,  
$G_{\pm\pm}(t,t^{\prime}) = \sum_k\,i\bigl(\langle T^{\pm}Q_k(q_k(t))
Q_k(q_k(t^{\prime})) \rangle - \langle Q _k(q_k(t)) \rangle \langle
Q_k(q_k(t^{\prime})) \rangle \bigr)$ and  
$G_{+-}(t,t^{\prime}) = G_{-+}(t^{\prime},t) = \sum_k \, -i \bigl(
\langle Q _k(q_k(t)) Q _k(q_k(t^{\prime })) \rangle - \langle
Q_k(q_k(t)) \rangle \langle Q _k(q_k(t^{\prime })) \rangle \bigr)$.
All the expectation values are considered with respect to the asymptotic
{\sl in} vacuum $|0\ \rm{in}\rangle_I$ in the interaction picture.
Note that we have integrated out the environment degrees of freedom and
$S_{IF}$ depends only on the system variables.

It is important to separate the real and imaginary parts of the influence
action because, as is well known \cite{feynman,calzetta94}, the imaginary
part is related to the noise that the environment induces on the system,
whereas the real part gives the averaged dynamics of the system. These
are:

\begin{eqnarray} 
\label{7}\Re S_{IF}[x_{+},x_{-}]&=&\sum_k \left[ \int dt\left\langle Q_k
\left(q_k\left(t\right) \right) \right\rangle \Delta \left( t\right)
+ \frac 12 \int dtdt^{\prime }\Sigma \left( t\right) H_k\left( t,t^{\prime
}\right) \Delta \left( t^{\prime }\right) \right] ,\\ 
\label{8}\Im S_{IF}[x_{+},x_{-}]&=&\sum_k \left[ \frac 12\int dtdt^{\prime
}\Delta \left(t\right) N_k\left( t,t^{\prime }\right) \Delta
\left(t^{\prime }\right) \right] , 
\end{eqnarray}
where we have defined $\Sigma \left(t\right) \equiv h\left( x_{+}\left(
t\right) \right) +h\left( x_{-}\left(t\right) \right) $ and $\Delta \left(
t\right) \equiv h\left( x_{+}\left(t\right) \right) -h\left( x_{-}\left(
t\right) \right)$ and we have introduced 

\begin{eqnarray} 
\label{9}
H_k(t,t^{\prime}) = A_k(t,t^{\prime})-D_k(t,t^{\prime}) =
-2D_k(t,t^{\prime}) \theta(t-t^{\prime}) ,
\end{eqnarray} 
which has been expressed in two alternative and equivalent ways for
further use. Here the kernels $A_k$, $D_k$ and $N_k$ are defined as
follows: $D_k(t,t^{\prime}) = -\frac i2\left\langle \left[ Q_k\left(
q_k\left(t\right) \right) ,Q _k\left( q_k\left( t^{\prime}\right) \right)
\right]\right\rangle $ is the dissipation kernel and $N_k(t,t^{\prime}) =
\frac 12\left\langle \left\{ Q_k\left( q_k(t)\right) ,Q _k\left(
q_k(t^{\prime}) \right) \right\} \right\rangle-\left\langle
Q_k\left(q_k(t) \right) \right\rangle \left\langle Q_k(q_k(t^{\prime}))
\right\rangle$ is the noise kernel. The dissipation and noise
kernels, which are related by the  fluctuation-dissipation theorem, are
antisymmetric and symmetric respectively under interchange of $t$ and
$t^{\prime}$. On the other hand, the kernel $A_k\left( t,t^{\prime
}\right) =\frac i2 \mathop{\rm sgn}(t-t^{\prime}) \left\langle \left[ Q
_k\left( q_k(t)\right) ,Q_k\left(q_k(t^{\prime}) \right) \right]
\right\rangle $
is symmetric and, as we will see, it is the part that gives rise to
divergences.
 
\section{SUM OVER ALL THE MODES AND NEED FOR RENORMALIZATION} 
\label{sect3}

For concreteness, let us now consider the case $Q _k\left(
q_k(t) \right) =\frac g2 q_k(t)^2$ ($g$ is a perturbative coupling
constant) where $\phi(t,\vec x)$ is a massless real scalar field
satisfying the Klein-Gordon equation in Minkowski space-time. We can use
the following conventions (note that the label for each mode, $k$,
corresponds in fact to a three dimensional vector):
$\hat{q}_k\left( t\right) =\hat{a}_k f_k(t)+\hat{a}^\dagger_k f_k^{*}(t)$,
$G_k^{+}(t,t^{\prime}) \equiv \langle\hat{q}_k(t)\hat{q}_k(t^{\prime})
\rangle = f_k(t)f_k^{*}(t^{\prime})$ and  
$G_k^F(t,t^{\prime}) \equiv \langle T\hat{q}_k(t) \hat{q}_k(t^{\prime})
\rangle =\theta(t-t^{\prime}) f_k(t)f_k^{*}(t^{\prime}) +
\theta(t^{\prime}-t) f_k(t^{\prime})f_k^{*}(t)$, 
where $\hat{a}^\dagger_k$ and $\hat{a}_k$ are the creation and
annihilation operators for each of the modes $u_k(\vec{x})$ in which the
field
$\phi(t,\vec x)$ has been decomposed. When properly normalized,
$f_k(t)=(2\pi)^{-\frac 32} (2 \omega_k)^{-\frac 12} \exp(-i\omega_kt)$
with $\omega_k=({\vec k}^2)^{\frac 12}$. Taking all this into account, we
will have
 
\begin{eqnarray} 
\label{15} 
D_k\left( t,t^{\prime }\right) &=& -
\frac i4\left[ G_k^{+}\left( t,t^{\prime }\right) ^2-G_k^{+}\left( 
t^{\prime},t\right) ^2\right] ,\\
A_k\left( t,t^{\prime }\right) &=&  
\frac i4 \mathop{\rm sgn} 
\left( t-t^{\prime }\right) \left[ G_k^{+}\left( t,t^{\prime }\right) 
^2-G_k^{+}\left( t^{\prime },t\right) ^2\right] ,\\
N_k\left( t,t^{\prime}\right) &=&\frac 14\left[ G_k^{+}\left( t,t^{\prime
}\right) ^2+G_k^{+}\left( t^{\prime },t\right) ^2\right] . 
\end{eqnarray} 
To perform the sum over all the modes, we note that we may write
$A_k+iN_k=\frac i2{G_k^F}^2$ and $D_k+iN_k=\frac i2 {G_k^{+}}^2$. Using
the integral representations for $G_k^F$ and $G_k^{+}$,  
$G_k^F(t,t^{\prime})=-(2\pi i)^{-1} \int_{-\infty }^\infty d\omega
e^{-i\omega (t-t^{\prime})}(\omega ^2-\vec k^2+i\varepsilon)^{-1}$,
$G_k^{+}(t,t^{\prime}) = \int_{-\infty }^\infty d\omega e^{-i\omega
(t-t^{\prime})} \delta (\omega ^2-\vec k^2) \theta(\omega)$
we obtain
\begin{eqnarray} 
\label{17} 
G_k^F\left( t,t^{\prime }\right) ^2&=&-\frac 1{\left( 2\pi \right) ^2}
\int_{-\infty }^\infty dk^0e^{-ik^0\left( t-t^{\prime }\right) }
\int_{-\infty }^\infty \frac{d\omega }{\left[ \omega ^2-\vec k^2 + i
\varepsilon \right] \left[ \left( \omega -k^0\right) ^2-\vec k^2 + i
\varepsilon \right] } ,\\
G_k^{+}\left( t,t^{\prime }\right) ^2&=& \int  
\nolimits_{-\infty }^\infty dk^0e^{-ik^0\left( t-t^{\prime }\right) } \int  
\nolimits_{-\infty }^\infty d\omega \delta \left( \omega ^2-\vec
k^2\right) \theta \left( \omega \right) \delta \left( \left( \omega
-k^0\right) ^2-\vec k^2\right) \theta \left( k^0-\omega \right) .
\end{eqnarray} 
When carrying out the sum over modes ( $\sum_k\equiv V/(2\pi)^{n-1}
\int d^{n-1}k$ ) we note that the $G_F^2$ term will
diverge. Thus, we use dimensional regularization to perform the integrals
and then expand in powers of $(n-4)$, where $n$ is the spacetime
dimension. The usual procedure gives:

\begin{eqnarray} 
\label{18} 
A\left( t-t^{\prime }\right) +iN\left( t-t^{\prime }\right) &=&-  
\frac V{32\pi ^2} \int  
\nolimits_{-\infty }^\infty dk^0e^{-ik^0\left( t-t^{\prime }\right) 
}\left[\frac 1{n-4}+\frac 12\ln \left( \frac{(k^{0})^2+i\varepsilon }{\mu 
^2}\right) \right] ,\\
\label{19}
D\left( t-t^{\prime }\right) +iN\left( t-t^{\prime }\right) &=& 
\frac{iV}{32\pi ^2} \int  
\nolimits_{-\infty }^\infty dk^0e^{-ik^0\left( t-t^{\prime }\right) 
}\theta \left( k^0\right) ,
\end{eqnarray} 
where $A=\sum_k A_k$, $D=\sum_k D_k$ and $N=\sum_k N_k$.
The second integral is finite and thus, we finally have the finite parts: 
$A_{ren}(t-t^{\prime}) = -(V/32\pi^2) \int_{-\infty }^\infty dk^0
e^{-ik^0(t-t^{\prime})} \ln (k^0 / \mu) ^2$,
$D(t-t^{\prime}) = i (V/32\pi ^2) \int_{-\infty }^\infty dk^0
e^{-ik^0(t-t^{\prime})} \mathop{\rm sgn}(k^0)$
and $N(t-t^{\prime}) = (V/32\pi) \delta (t-t^{\prime})$.
The divergent part $A_{div}\left( t-t^{\prime }\right) =-(V/16\pi (n-4))
\delta (t-t^{\prime})$ has been separated in such a
way that the divergences may be absorbed by counterterms in $S[x]$. In
other QFT contexts (e.g. two interacting scalar fields)
\cite{hu93,calzetta97} the finite contribution from the counterterms can
be reabsorbed in the renormalized parameters. However, as we will see, in
semiclassical gravity some logarithmic finite terms which cannot be
reabsorbed arise in the counterterms.
 
\section{STOCHASTIC SEMICLASSICAL GRAVITY} 
\label{sect4}

As an example we consider the back reaction due to the effect of a small
mass or a non-conformal coupling of the scalar field $\phi(x)$ on a flat
Robertson-Walker model \cite{calzetta94,hu95,calzetta98}. We have to make
the following substitutions:

\begin{equation} 
\begin{array}{c} 
x\left( t\right) \rightarrow a\left( \eta \right) \\  
Q _k\left( q_k\left( t\right) \right) \cdot h(x(t)) \rightarrow \frac
12 \phi _k(\eta) ^2\Delta \omega^2\left(a(\eta)\right) =\frac 12 \phi
_k\left( \eta\right) ^2\left( m^2+\left( \xi -\xi _c\right) R\left(a\left(
\eta \right)\right) \right) a\left( \eta \right) ^2 ,
\end{array}
\end{equation} 
where $\eta$ is the conformal time, $a(\eta)$ the scale factor,
$R(a(\eta))$ the scalar curvature, $m$ the scalar field mass and $\xi$ a
dimensionless constant. In those previous works where Einstein-Langevin
equations were derived using mode decomposition, divergences were not
dealt with \cite{calzetta94,hu95}.
 
Let us now see how special care is needed with the sum of modes. Take for
instance the second definition for $H_k$ in (\ref{9}) and note that, using
the real part of (\ref{19}), $D(\eta-\eta^{\prime})=\sum_k
D_k(\eta-\eta^{\prime})$
may be written as $(V/16\pi^2)\,\mathop{\rm PV} (1/(\eta
-\eta^{\prime}))$. In this case, one would be inclined to write
$H=\sum_k H_k=D(\eta,\eta^{\prime}) \cdot \theta(\eta-\eta^{\prime}) =
(V/16\pi^2)\,{\rm PV}(1/(\eta -\eta^{\prime})) \cdot
\theta(\eta-\eta^{\prime})$, 
but this is an ill-defined product of distributions which may give rise to 
divergences. A possible way to deal with this is by using, instead, the
first definition in (\ref{9}) and consider $A$ and $D$ separately:
\begin{eqnarray} 
\label{20}
H=\sum\limits_kH_k=\sum\limits_kA_k-\sum\limits_kD_k ,
\end{eqnarray} 
where the first term in the last member will be ultraviolet divergent
whereas the last term is finite. Now the divergence can be clearly
identified and one may use the proper counterterm in dimensional
regularization to cancel it: 
\begin{eqnarray} 
S_g^{div} [a(\eta)] &=&  
\frac{\left( \xi -\xi _c\right) ^2\mu ^{n-4}}{32\pi ^2\left( n-4\right) }% 
\int d^nx\sqrt{-g}R^2=\frac{\left( \xi -\xi _c\right) ^2}{32\pi ^2\left( 
n-4\right) }V\int d^{n-3}x\left[ \frac{36}{n-4}\left( \frac{\ddot{a}}{a}% 
\right) ^2\right. \nonumber \\ 
&&\left. +36\left( \frac{\ddot{a}}{a}\right) \left( \ln 
\left( a\mu \right) \left( \frac{\ddot{a}}{a}\right) +\frac 23\left( 
\frac{% 
\dot{a}}{a}\right) +\left( \frac{\dot{a}}{a}\right) ^2\right) \right] . 
\end{eqnarray} 
The second term in this integral, which is finite, will cause the
appearance of extra terms when deriving the Einstein-Langevin equation.
Using now the results of the previous section, we get total agreement with
those results reached by functional methods which do not use mode
separation \cite{calzetta98}. 
 
A very interesting connection between dissipation and fluctuations in the
metric and particle creation has been revealed by Calzetta and Hu
\cite{calzetta94}. They computed the energy dissipated by the
gravitational field per unit volume as $\rho _d = \int d\eta
d\eta^{\prime} (\partial H(\eta,\eta^{\prime})/\partial \eta) \Delta
\omega^2 (\eta) \Delta \omega^2 (\eta^{\prime})$ (for simplicity we have
considered that the asymptotic values of the scale factor are $a_{in} =
a_{out} = 1$), and showed that it was equal to the energy density of the
created particles,
$\rho _{created \atop particles} = (2\pi)^{-3}\int 4\pi \nu ^2V\nu |\beta
_\nu| ^2d\nu$ , where $\beta _\nu$ is the Bogoliubov coefficient for the
modes with frequency $\nu$.
However, formal use of divergent expressions was made in such a
derivation. Our treatment shows clearly that the divergent part $A(\eta,
\eta^{\prime})$ of the kernel $H(\eta ,\eta^{\prime})$ decomposed
according to (\ref{20}) gives no contribution since it is symmetric under
interchange of $\eta$ and $\eta^{\prime}$ and hence the derivative will be
antisymmetric:

\begin{eqnarray}
\rho _d  &=& \int d\eta d\eta^{\prime} \frac{\partial H\left(
\eta,\eta^{\prime} \right) }{\partial \eta }\Delta \omega^2 \left( \eta
\right) \Delta \omega^2 \left( \eta^{\prime} \right) =-\int d\eta
d\eta^{\prime} \frac{\partial D\left( \eta ,\eta^{\prime} \right)
}{\partial \eta}\Delta \omega^2 \left( \eta \right) \Delta \omega^2 \left( 
\eta^{\prime} \right) .
\end{eqnarray} 
This integral is, therefore, manifestly finite and can be computed using
the dissipation kernel obtained from eq. (\ref{19}) thus leading to the
same result of Ref. \cite{calzetta94} without need to deal with divergent
expressions.

\section*{ACKNOWLEDGMENTS}

It is a pleasure to thank Esteban Calzetta and Rosario Mart\'\i n for
useful comments and stimulating discussions. This work has been partially
supported by the CICYT Research Project number AEN98-0431. A.R. also
acknowledges support of a grant from the Generalitat de Catalunya.

\end{document}